# When is now in a distributed system?

# Animated motion (could) set the present in brain networks

Julien Lagarde[1], Nicolas Bouisset[2]

[1] Euromov, Montpellier University France

[2] Department of Kinesiology, Western University, London, ON, Canada

Acknowledgements:

This study was financially supported by the European Project ENTIMEMENT, H2020-FETPROACT-2018, Grant Number 824160. The authors thank Anders Ledberg for insightful discussions, and Phil Fink, JL Vercher, and G Zelic, for technical help with the experiment.

**Abstract:** Our brains are viewed as interconnected distributed systems. The connections between distant areas in the brain are significantly delayed. How to obtained *now* in such networks with delayed interconnections? We aim to show that delayed communication and interconnectedness of the brain impose an interaction with the environment, assuming that such an access to *now*, which we label *t-present*, is of use for this system. It is conjectured that for any sensory, motor or cognitive functions to work efficiently an updated sort of "time origin" is required, and we claim that it is uniquely given by a direct contact with the physical environment. To get such contact autonomously any movement is required, be it originating in the motion of sensory systems or in goal directed movements. Some limit cases are identified and discussed. Next, several testable situations are envisioned and available studies in favor of the main theoretical hypothesis are shortly reviewed. Finally, as a proof of concept, an experimental study employing galvanic vestibular stimulation is presented and discussed.

That physical actions performed by athletes, artists, artisans, surgeons, pilots, are fine tuned to the environment is in plain view to the spectator and duly confirmed by careful measurements. Indeed spatial and temporal coincidence is accurately achieved up to an extreme degree, and the efficient adjustment of forces seems also boundless. The timing of coordinated movement defies the spectator's eye, and perceptual sensitivity can evolve to grasp surprisingly small details. Such an exquisite tuning to the environment makes a radical embodiment stance on the underlying functions natural, and already proves fruitful among the various alternative theoretical approaches currently available.

From a bird's eye, this fine tuning arises at different temporal scales by trial and error, through natural selection, development, learning, and culture, enabling acquisition and individual memorizing and inter-personal - cultural archiving of skills to produce the intended effects in the world. For (goal directed) acting in the world, the necessity of a coupling to the environment can hardly be discussed. Still, finding how to formulate it to discover which laws govern it's functioning, to reach predictive and generalization power, is a great challenge.

In line with the manifold theoretical roots of the so called embodiment approach (Clark 1998; Kelso, 1995; Turvey et al. 1981; Varela et al., 1991) we assume that this coupling to the environment have recurrent consequences on elementary functions, not restricted to highly skilled behavior, or to the development and initial physical grounding of agency or of cognitive functions.

The present paper presents a theoretical and speculative proposal, which we suggest contribute to the study of radical embodiment of sensorimotor skills and other general functions. After outlining the main ingredients of our theoretical hypothesis, we then envision several options for operational testing. We first present a theoretical, though speculative, argument for the necessity for animated motion, basically the action in a physical world, for proper brain functioning. Specifically, it is conjectured that one pervasive organization of the brain prevents autonomy from the action in environment: The presence of delays of interaction between its connected elements. If solid enough, the interest of the proposal is that it comes as a consequence of inner properties of the brain dynamics, which seems as a logical twist with respect to the emblematic quote from Bill Mace's, an influential thinker of the ecological psychology inspired by James Gibson: "Ask not what's inside your head, but what your head's inside of". Here we suggest that the analysis of dynamics arising inside of the head call for an ecological neuroscience (See for a similar detour, but in a distinct context Salinas, 2006).

**Statement of the requirements: A toy model three nodes ring with delays**

Delays are ubiquitous in the brain, given the length of connections and fixed conduction velocity. To be more specific in the most standard way delays are modeled by delayed differential equations (Equation 1):

$$dX/dt = F(X(t), β) + G(X(t-tau), λ) \qquad (1)$$

The derivative of the current state is a function, eventually taking parameters β and λ, F of the current state X(t) and G of the state at a past time t- tau X(t-tau), tau being the delay. This equation is shown here only to provide an explicit illustration of the concept of time delay. By virtue of infinitesimal calculus, delayed dynamical systems are infinite dimensional. The current state derivative is a function of the whole interval between current time (t) and the past time (t-tau). Time steps being infinitely smalls, an infinite number of intermediate past states affect the left hand side derivative. Note that the numerical integration of such equations is not trivial (Shampine & Thompson, 2001). In the following we won't deal with the (difficult) study of such dynamical systems (see a recent example in Zheng & Pikovsky, 2019, and references therein).

One can cite multisensory and sensorimotor integration as an example of function involving brain networks with significant delays, supported by large scale networks between distant areas (Barnejee & Jirsa, 2007; Beuter et al., 1993; Chen et al., 1997; Milton et al., 2008; Tass et al., 1997; Thakur et al., 2016; Venkadesan et al., 2007; to name a few), but this property is ubiquitous in large scales brain networks. This property has been studied for its effect onto dynamics: Destabilization, oscillation onset, multistability, 1/f fluctuations.

Consider now as a toy model a three nodes case with three delayed coupling (Figure 1 and Equation (2)). Each node evolves with its own dynamics, steady state or oscillatory for example, which will not be considered here. Our toy model as the structure of a ring, which may highlight most evidently the property of interest here. Such a structure for networks has been often studied (Collins & Stewart, 1994), with application to central patterns networks, hence oscillating nodes, notably. Rings of unidirectional delay coupled excitable nodes have also been analyzed recently (Zheng & Pikovsky, 2019). Rings of neural ensembles are pervasive in the nervous system and widely used in modeling, illustrated by closed loop between distributed areas, linking for instance basal ganglia, cortex and cerebellum, or thalamus and cortical distributed areas (Sporns et al., 1989; Cappe et al., 2009). The description here is kept on purpose at it utmost simplicity, and the reasoning below is by no way considered as a proof but rather as a theoretical proposal, or a conjecture.

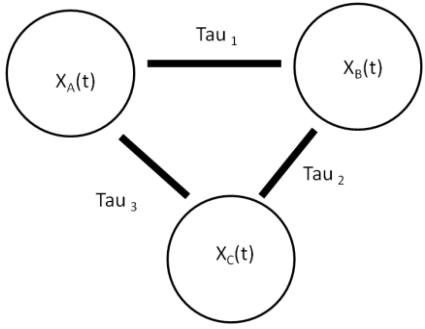

Figure 1- An illustration of an interconnected system. Is shown a ring composed of three nodes A, B, and C, each one with current states $X_i(t)$, coupled with time delays $tau_{(i)}$ = 1, 2, 3. For the present purpose considering the simplest case is sufficient, that is, the nodes and coupling are equivalent, in other words they are said invariant by permutation.

$$dXA/dt = F(XA(t)) + G(XB(t-tau_1), XC(t-tau_3))$$

$$dXB/dt = F(XB(t)) + G(XA(t-tau_1), XC(t-tau_2))$$

$$dXC/dt = F(XC(t)) + G(XA(t-tau_3), XB(t-tau_2)) \qquad (2)$$

Equations (2) represent the 3 nodes ring of Figure 1, parameters are dropped for clarity. In this case the couplings are bidirectional or reciprocal, thus B&C act on A, A&C act on B, A&B act on C. The delays, taus, can be or not equal. It is noteworthy that change from birectional to unidirectional have drastic effects on the dynamical behavior of rings (Zheng & Pikosky, 2019), but the behavior of (2) or of possible variants will not be addressed in this paper, our reasoning will be only conceptual and not mathematical. In Thakur et al. (2016) a 3 nodes system has been studied, each node representing a neurons population using a Kuramoto oscillator, 2 nodes are reciprocally coupled with delays while the third is coupled without delays with one of the first.

**A thought experiment: Introducing the *present*.**

From our 3 nodes toy model with delays, let's attempt a though experiment. To emphasize the consequences of time delays in this loop, we assume that this 3 nodes system can be mapped onto another three nodes system with one, say A(Xa(t)) being functioning in a timeframe that we call *t present* while the 2 others, B and C, are functioning in past timeframes. We have change the time origin for the different nodes. After this transformation it can be considered that the coupling is no longer delayed but instantaneous (Figure 2). In this case one may imagine that two nodes B and C are "observed" by the one in the *t present*, A. A could also be considered as marking the time origin, thus relative to it, that is relative to the information A could receive, B and C are functioning in the past. Let's imagine a sort of time travel experience. Picture yourself as being A, you are receiving at each instant information about the state of B and C coming from their past, like the light of a long dead star reaching the telescope on the earth surface. However one could stand also in place of B and be equivalently receiving information from the past of A and C. The same applies when standing in the C spot on the ring.

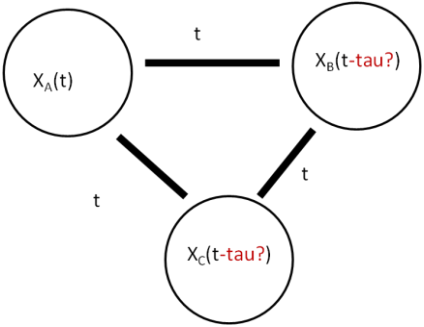

Figure 2- The 3 nodes ring model after shifting the timeframe origins. In this representation the couplings are instantaneous but each node functions according to distinct timeframes origins. Here the node A($X_A(t)$) is taken as the *present*. Note that the interrogation along with the delay value tau "tau?" in B and C indicates clearly that the authors are not aware of a proof of equivalence between the two representations of the delayed ring.

Consider this way, one can see that in this ring model, nodes can be exchanged and timeframe origin cannot be established in a unique way in the model itself. Timeframe alignment is so to speak degenerate, or "ambiguous". However open to criticisms, the present thought experiment proposes that we have now a system with three entities functioning at distinct times. This emphasizes that if for proper functioning it is required to set what one calls basically the "present" in the ring network, then this has to be done from outside of the ring system. Simply put: Present can't be internally obtained. One may think that the interaction with a supplementary network could settle this issue. It is common practice in brain network modeling to set up distinct sub networks, for example to dispatch memory persistence in one network and read out of this memory in another network (See Seung, 1996). We assume also that this will not solve the problem, and that an initial condition will always be required to get a mark of a timeframe origin. This dependence on the initial state has been identified as a current limit to, dynamical systems, state-dependent networks models of temporal processing (Karmarkar & Buonomano, 2007) (See the Annex 1). We suggest that a parsimonious solution to find ground for "present" requires to get out, hence to get contact with the physical environment. And we conjecture further that there no other way around, therefore this would be a necessity. If the present argument is solid, the need to tag networks dynamics to outside events makes movement in the physical environment mandatory.

It is noteworthy that reference is often made in the present paper to models of so called temporal processing. We suggest that our conjectural proposition is not restricted to this functional context, the temporal processing, in which implicitly or explicitly time has to be dealt with at the level of the function/ behavior. Indeed the tasks used to address temporal processing are not key in principle here, the effect of delays on network we focus on is presumably pervasive to a large if not all, brain networks. However, those temporal tasks may be used as practical and natural measurement procedures to investigate which factors perturbs or ease the access to *t - present* in the brain. Tasks like duration judgment, order judgment, causal judgment, rhythmic movements self-paced, or synchronized to external events can be taken as a straightforward entry point, while still ambitioning that the conclusions are general.

**"Embodied movement devices" to get *t- present***

Above we assumed that *t-present* cannot be obtained in interconnected neurons populations, and that this timeframe origin is useful for proper brain functioning. We propose that this *t-present* can be obtained in action-perception loops, that is each time occurs a relation between movement and its causal perturbation of sensory systems. Let us review in the following some examples of what we dubbed "devices" to get *t-present*. In a reaching movement to an object the physical touch with the object can correspond to the event that provides a *t-present* milestone. The same time milestones are available during upright standing, or even while seating on a chair, as small but permanent motion of the head and torso, or legs, can be sensed via different sensory systems. In the postural case one can cite of course among others the vestibular system which gets access to acceleration due to the gravitational field, but also motion consequences in optical flow or acoustic flow can be picked up, or muscle stretching in the torso, hips, and the upper limbs by and large including- legs-ankle-feet. Goal directed and the so called postural movements can afford this identifiable event, but even simpler devices, like tonic reflex against gravity may be sufficient to signal a contact with the outer world.

Example above are simple perturbations of the sensory systems by body motion happening at specific times, that permanently of intermittently recur to provide a new initial condition which separates past and future. Limiting cases here are perturbations of the senses which are not self-produced but passive, including any sense, a light, a sound, a tactile contact, a passive limb movement, or an odor. Yet another relevant distinction may concern attention orienting and whether such a passive event is expected or not. A second limiting case could be a touch event self-produced, like touching one hand with the other. At first sight this is not literally a contact with the outer world but a body self-contact.

Furthermore, one can think next to an access to a physical present caused by self-motion in relation to a coupling body-outer world not so much parallel or incidental but directly relevant to a behavioral function and goal. We can cite all the synchronization involving body-world coupling, including rhythmic or discrete cases. It is noteworthy that such coupling to cause, sense and adjust to outer world was suggested to give support to the emergence of agency in early development (Kelso & Fuchs, 2006). Interestingly, the dynamical model of rhythmic synchronization applied to sensorimotor synchronization, basically a self-sustained rhythmic movement being entrained to an external driving rhythm, is dubbed non autonomous: There is a differential equation which is a function of time. That is, time is explicit in the periodic forcing function, say $A \times (\cos(\omega \times t))$, that represents the (sensory) perturbation by the external event, on the left hand side of the dynamical system. In a distinct category one can cite the family of avoidance and control laws studied after David Lee's time to contact "tau" (Lee, 1976; see Schöner & Dose, 1995; Warren, 2006), in which movement is coupled to an obstacle or to the trajectory of an object. In that respect an interesting candidate device can be to act to meet the time left to avoid falling on the ground; that limit time may set the time landmark we are seeking to get our brains inner dynamics framed. No matter which coordination pattern is used, which varies between individuals depending on muscle strength, proprioception, body scale and the like, what may count is that a coupling to resolve a physical constraint is set in, that it creates an event which again have a date. Hence a broad class of adjusting posture to avoid falling, or making a step to avoid it, requires attunement, that is a proper adaptation, an adjustment of time scale, to the physics of falling objects, which implies gravitational force and acceleration. To elaborate further, the last example may provide an absolute "metric", or unit, of time duration to the brain. What is the range and average time to fall given your size and height of your center of mass? In the same vein aiming to catch a falling object requires reference of movement to a physical event in motion independently to the body movement. In the same vein, consider an event caused (or not) by movement which stimulates different sensory systems operating at different speeds and communicating to others parts of the brain with different time delays. For a given event, physiological consequences of sensory perturbations in distinct modalities by physical change are not time aligned, say in a given context the brain faces an average delay of "N" ms between any pair of senses, this may provide also an absolute metric – a unit- for durations.

We briefly reviewed examples above showing that several redundant origin of time basis could be obtained, providing our *t-present*. Such encounters with "present" can be regularly updated, estimated, and possibly transiently sustained with the help of memory. Accordingly, time origin maybe probed…from time to time.

**This "conjecture" needs predictions**

We assumed above that "being there" (Clark, 1998) is important for brain functions, and requires recurrent, permanent or intermittent, probing interaction body-world motion events. Is this hypothesis testable? One has to prove that (i) getting t-present is important, (ii) some of the devices proposed above are used to get t-present at the exclusion of others, and (iii) in the long run getting t-present can be related to pathology. Related to the last point one message of radical embodiment, or ecological psychology, remember Bill Mace's advice, is to pay prior attention to the relation of the individual to his environment. In relation to the so called motor disorders, motor output, aka motor planning, is very often under scrutiny while perception is so to speak occluded. Much has to be done in that domain attempting instead to restore perception –action coupling, or loops in another scientific jargon. A recent review echoed this line of thinking for Parkinson disease, advocating that disrupted tactile and proprioceptive sensations alter "interaction between sensory input and motor output" (Conte et al., 2013).

Maybe the main prediction we come across is that brain networks involved in or "sensing" this access to time origin must be connected to any other network so as to spread the time origin resetting. Second, one may try to perturb the access to the time origin and check the consequences on given functions, temporal ones of course, like continuation task in simple movement tapping, or self-paced movement, but as suggested anyone should be impacted. The difficulty lies in perturbing in a way which is complete, that is covering the multiples sources, as we saw time origin devices are manifold. Perturbations of the vestibular system using GVS (galvanic vestibular stimulation), or virtual reality experiments, could be a venue to suppress or disturb time origin probes. As a proof of concept we will present preliminary experimental evidence of the effect of GVS onto rhythmic movements in the next section. Some existing data already give flesh to the assumptions presented here. Firstly it was shown that microgravity, or self-motion, can affect several timing functions (Binetti et al., 2010; Capelli et al., 2007; Dallal et al., 2015; Jörges & López-Moliner, 2017; Semjen et al., 1998; Moscatelli

& Lacquaniti, 2011). However, by performing the task itself the brain has access to t-present, locally at the moving limbs, hence the vestibular perturbations may not be sufficient, unless the vestibular system plays a central role. Furthermore, the perturbation used may be non-specific. It is difficult to exclude the possibility that the modulatory effects found are the consequence of a perturbation of the brain, for instance of attentional processes, or of multisensory integration processes, but not specifically of the devices accessing t-present. It is noteworthy that both limitations apply also to our own experimental study presented next.

Secondly, psychophysical findings indicate that subjective time duration judgment is flexible. Duration judgments are distorted during slow-motion video sequences of natural biological motion (Eagleman, 2004). Furthermore, when we attribute the cause of events to our actions, it is perceived as occurring earlier (Haggard et al., 2002). Finally, perceived temporal order of action and sensation can reverse after an adaptive recalibration process (Stetson et al., 2005).

**Experimental proof of concept**

Aims

Based on our assumptions, and the previous studies shortly reviewed above, we predicted that GVS will perturb the performance of simple index finger movements. We performed an experiment on the coordination of rhythmic index movements in anti- phase (Kelso, 1984), with and without GVS. The variability of periods, and of maxima of flexions and of extensions was estimated. The participants were asked to synchronize their right index finger with a beat. A control experiment was performed with a subset of the participants (N = 6), in a self-paced condition, without a metronome. The second experiment is not reported, as no significant effect of the GVS was found. Those data were collected as controls in a broader project for the Master Dissertation of the second Author. Here we report only control conditions with without GVS, however due to the overall project the controls were not completely natural, participants were asked to look at the left index.

Participants

All participants were right handed and were totally naïve to the purpose of the study. They all had normal or corrected-to-normal vision and no history of musculoskeletal problems, vestibular or neurological disorders. Ten healthy volunteers (two females, mean age 25 years old ± 3 years) participated in the experiment. Written informed consent was obtained from each participant prior to the experiment. The study was conducted in accordance with the Declaration of Helsinki, and approved by the local ethics committee of the University of Montpellier, France. All participants gave written informed consent before inclusion.

Apparatus

The auditory stimuli were 80 ms square wave pulses with a tone carrier frequency of 500 Hz. The experiments were conducted using two PCs, one devoted to stimulus presentation, the other used to record simultaneously fingers movement and stimuli, via an A/D card (NI USB- 6009, National Instruments). The stimuli, controlled using the data acquisition toolbox Matlab (Mathworks), were sent via the sound card to a hardware system (Arduino 1.0.5). This device was used to deliver auditory stimuli while avoiding electronic delays. Index finger positions were recorded using two types of electro-goniometers (SG65 Biometrics Ltd, resolution ± 2°, for the first experiment and resistive flex sensor; Spectra Symbol; resolution: 1 degree for the second and the third experiments). These electro-goniometers were attached to the back of the hand to estimate the metacarpophalangeal joint position of the index fingers.

Fingers' angular positions and stimuli were collected by a second PC at a sampling frequency of 500 Hz in the first experiment and at 3000 Hz in the two-other experiments using an A/D NI USB acquisition board (6009). Recording was controlled by a custom program using the functions of the data acquisition toolbox Matlab (The authors thank Phil Fink).

Procedure

Participants were asked to sit comfortably, feet together flat on the floor, with their forearms placed horizontally on a table. Their trunk was slightly bent for their head to be vertically placed relative to their fingers. They were asked to maintain their head as stable as possible. Participants were instructed to perform oscillatory movements, about the metacarpophalangeal joint, with their index fingers moving in an anti-symmetrical pattern at a comfortable amplitude. In such pattern, the flexion of a finger is performed simultaneously with the extension of the other one. Thus, while the right finger is flexed, the left one is extended and vice versa. In addition, for the first two experiments, the participants had to synchronize the movements with the external metronome by matching the flexion of the right index with the metronome's beep. During the third experiment, the participants no longer heard the metronome that was only given to the experimenter through headphones. The experimenter indicated the onset of each frequency step by taping on the desk with a pen. The participants that already participated in the previous experiment were trained during five minutes with the task without the metronome's beep. A ten-minute training period was given to the participant with no previous experience.

The metronome's frequency was 1Hz. They were instructed to stay with the metronome's tempo. At all times the participants were instructed to look to their left index finger. GVS was applied for half of the trials.

Galvanic vestibular stimulation (GVS)

A GVS stimulator (A395R Linear stimulus isolator, World precision instruments, INC Sarasota, FL, USA) applied a direct current to the subjects using a bipolar, binaural configuration, via electrodes placed over the mastoid processes. Electrodes consisted of customized metal electrodes (2- 3 cm) embedded in a sponge saturated with salted water to ensure proper conduction between the electrodes and the skin. Electrodes were fixed to the participant's head using kinesio-taping tape. The anode was placed behind the left mastoid process and the cathode behind the right mastoid process. The authors thank JL Vercher was enabling us to use this device. This montage is known to increase the vestibular firing rate on the cathodal side and decrease it on the anodal side (Fitzpatrick and Day, 2004). This mimic an inhibition of the left ear and an activation of the right ear. This montage was chosen for several reasons. It is known to activate predominantly the right hemisphere where the left hand is represented and there is growing evidence for the specialized role it plays in space construction mechanisms (Dieterich and Brandt, 2015; Karnath and Dieterich, 2006; Lopez et al., 2012).

A constant current of 1.5 mA lasting 1 min was reached after a 1 second ramp up progressive increase in stimulus intensity. At the end of the stimulation, intensity was decreased to 0mA after a 1 second ramp down. A 30 sec resting period was given between trials to dissipate the vestibular stimulation effects and to make sure the system reached its normal resting firing rate in between.

To ensure that the electrodes were well positioned, GVS was tested on postural control with their feet together and their eyes closed prior to data recording. In response to the GVS, subjects swayed towards the anodal side as reported in the literature (Fitzpatrick and Day, 2004; Guerraz and Day, 2005; St George and Fitzpatrick, 2011).

Data processing

Angular positions where low pass filtered (cut off 5 Hz), normalized and centered. The minima and maxima were identified and linearly detrended. For each trial 13 cycles were analyzed, thus 13 x 3 X 13 = 351 cycles were analyzed in each condition. The coefficient of variation was estimated for the periods, and the standard deviation was estimated for the minima and maxima of both index finger (see Figure 3 for illustration). The angular dispersion of the relative phase was estimated using a pointwise method between the maxima of flexion. The authors thank Gregory Zelic for sharing some programs for this analysis.

Results

Figure 3 show a sample trial. The positions in the middle panel column shows the strategy adopted by several participants: To synchronize the right index flexion on the beat and the left index extension on the beat (see Figure 3 captions for more details). This behavior resembles the slow and fast stages within each cycle of an excitable dynamics: Slow then fast to coincide with the beat (see Jirsa & Kelso, 2005; see Figure 6).

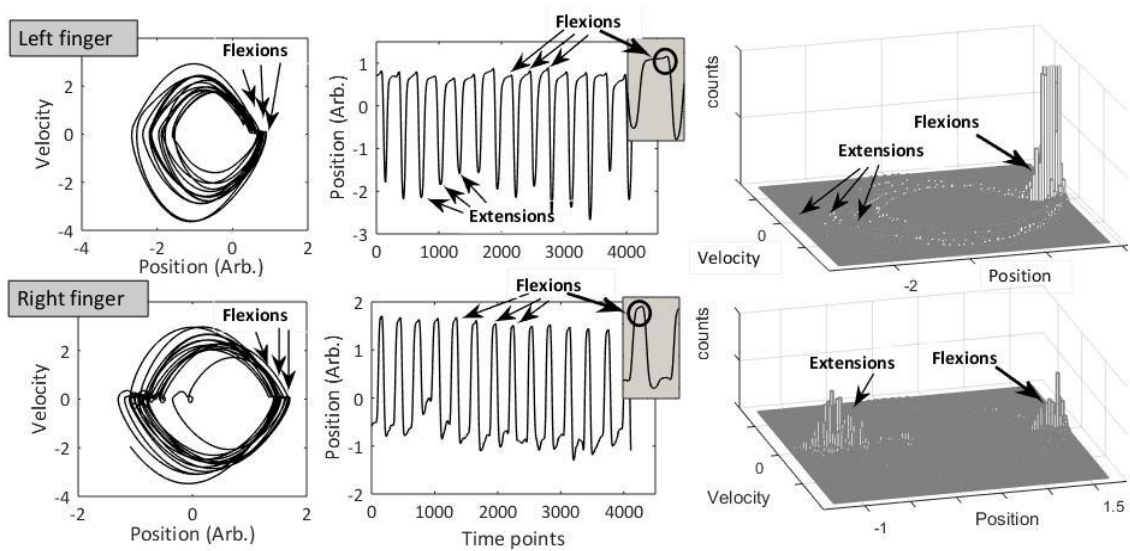

Figure- 3 One trial sample is shown to illustrate behavior and measurements. Top row: Left finger, lower row: Right finger. First panel: Phase portraits, middle: Position(t), and Right panel: Histograms of the phase portrait. Positive values of position correspond to maxima of flexion; Negative values of position correspond to maxima of extension. For each cycle: This participant slows down in extension with the right index, and with the left index slows down in flexion, then flexes fast to the beat with the right index and extends fast to the beat with the left index. The slow portions of the trajectories are clearly seen in the histograms on the right panels, where data accumulate with a small velocity at both extrema of position for the right finger, and at the flexions only for the left finger. Variability as a function of GVS was estimated at all extrema. The GVS was found to increase the variability at the maxima of flexion of the Left index, which is located at the end of the slow "resting" movement, right before a new extension to the beat is performed (see insert in the top graph in the middle column).

We performed Wilcoxon signed rank test on the variables including the 3 trials, thus 27 values per conditions. The GVs was found to increase the spatial variability at the maxima of flexion for the Left index ($p < 0.05$, $p = 0.03$). Without the GVS this variability was 0.048 (std 0.19) and with the GVS was 0.075 (std 0.059). A non-parametric t test using permutation confirmed the significant difference ($p = 0.023$; effect size = 0.66). One can question the particular sensitivity of the variable for which the variability is found to be affected by the GVS that would meet the $p<0.05$ threshold for decision, in comparison to other portions for which variability was analyzed, or the small sample. However, while inspecting thoroughly the data distributions, along the entire phase portrait trajectory (Figure 5), there were no other indication of a tendency for a variation due to GVS.

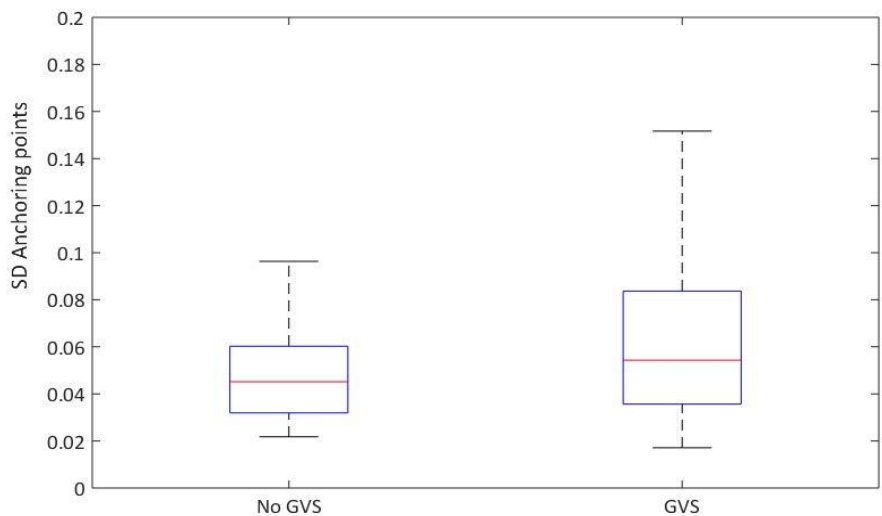

Figure 4. For the left index, without GVS and with GVS, the variability of the return position to flexion after an extension to the beat is shown. This return corresponds to the maximal value of the position for each cycle, that is the flexion (see Figure 3, top graph in the middle panel). For the trial shown in Figure 3 the arbitrary unit values of those maximal angular positions are centered around 1.5. The central mark in the box is the median, edges are the 25th and 75th percentiles, and the whiskers are drawn to the most extreme data points.

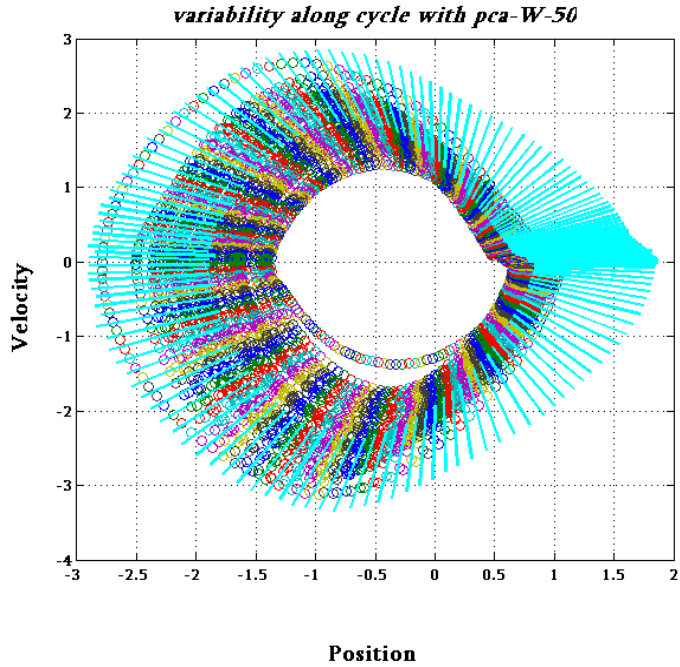

Figure 5. An example of the thorough inspection of the variability of the trajectories. Using local coordinates similar to a Frénet basis, calculated using principal components analysis (PCA) in sectors of the phase portrait, variance along the perpendicular (shown in blue) axis were analyzed.

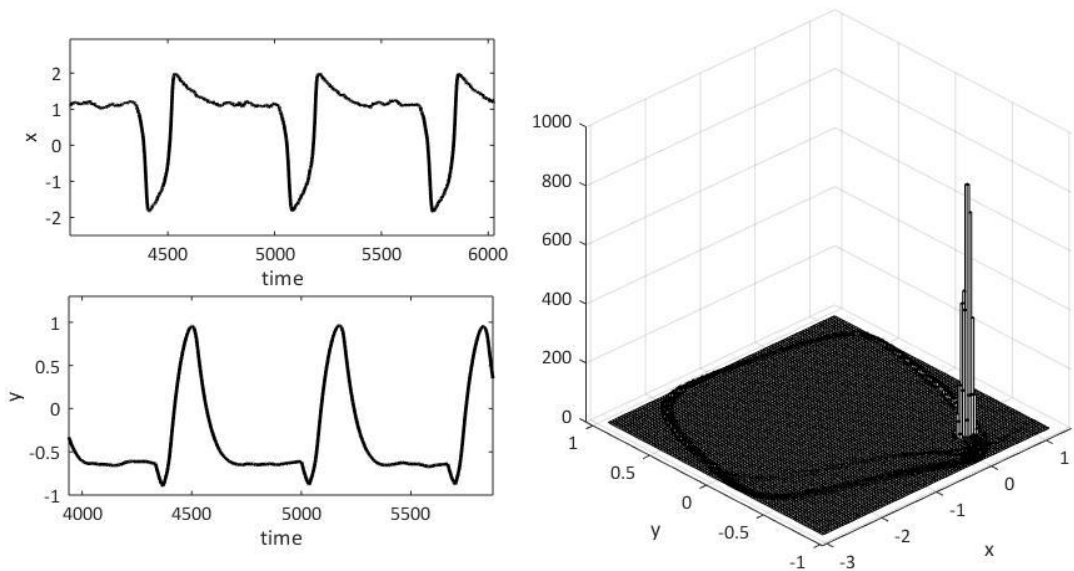

Figure 6. A numerical example of excitable dynamics time series similar to the behavior analyzed in the study (See Annex 2). The simulation of a single "excitator" with noise is shown (Jirsa & Kelso, 2005), driven by a series of discrete perturbations. Note that the "excitator" is not coupled to a second unit, like is to be expected in the experimental bimanual case, and there are no delays

involved here. We simply aim to show that dynamics resembling the data can be obtained with a low dimensional dynamical system (Ramdani et al., 2000). The slow evolution in the two coordinates labelled Y and Y is evident, followed by a fast and ample excursion, which could represent a flexion or an extension movement. One can also readily see overshoots and slow portions of trajectories, resembling the experimental data shown in Figure 3.

Discussion

The data collected have shown an effect of GVS on the variability of movement. This effect is present when participants were asked to synchronize their anti-phase pattern to a beat, but not when acting self-paced. Recall that synchronizing with a metronome belongs to non-autonomous dynamics, while self-paced belongs to autonomous dynamics. Hence we suggest that GVS has perturbed the coupling to the time dependent metronome. It is beyond the scope of this paper to analyzed in full details the consequences of our proof of concept. However, we may point at several issues. Firstly, it is evident that this experiment should be replicated with a single finger synchronizing to a beat to tackle more directly our assumptions. Secondly, to date, we can't provide an explanation of the increase of variability specifically at the slow dynamics location, and for the left finger. Variability is likely to be increased by a decrease of stability (attraction) of the slow motion trajectory. Clearly we cannot conclude that this is indicative of a reduced stability of the processes involved in the brain to access t-present. One may speculate that the slow (manifold) trajectory portion provides a time process regulating the preparation of the fast flexion or extension, before the occurrence of the beat, or regulating the period reproduction. Such a regulation could be performed by a state dependent process (Karmarkar & Buanomano, 2007), recurrently reset by access to the t-present.

However, it is not clear yet whether this effect proves the involvement of the vestibular system specifically to probe the outer world to get the t-present. Other interpretations are in order. As stated above, the GVS protocol may introduce an undifferentiated perturbation, hence the effect obtained could be caused by a process perturbed by the vestibular system, not specific to the hypothesized access to t-present. One candidate of such general cause mediating the increase of variability could be attentional orienting, known to significantly impact the coordination of movement (Mono et al., 2000). Another possible candidate is the perturbation of multisensory binding (Lagarde & Kelso, 2006), in which the vestibular system, including the so called vestibular cortex, is almost surely involved (Cullen, 2019). Nevertheless, the results obtained are in favor to the hypothesis that the vestibular system is active while coordinating the index fingers, but whether this indicates an access to t-present remains to be confirmed.

**Conclusion**

In the present paper we outlined a thought experiment to suggest that (many) brain networks may be ignoring which local activity is at present, in other words which local activity is past or future, eventually making any sequential ordering difficult. To do so we outlined the premises for a conjecture, which stands currently indeed speculative, having no rigorous proof, but that may be considered constructive. Other structure of interconnected networks, for instance line attractors, or forward networks, or networks with central relays, to name a few, may not suffer this weakness. We suggest however a fruitful program research which could investigate systematically this issue, and determine firmly the conditions to prove or disprove the conjecture and its potential consequences. Further we identified limit cases, which could dispute some tenets of the so called embodiment hypothesis, by which no contact to the outer world would be necessary, like self-touch, or passive perception. Next we proceeded to draw predictions for experimental testing. It is noteworthy that an experimental test can never replace a strong theory, but our preliminary experimental proof of concept is at least encouraging. To conclude, the analysis of limits imposed by the brain structure is a very intriguing venue to think about radical embodiment, considering an alliance between ecological psychology and neuroscience, for a lawful understanding of skillful behavior and general cognitive functions.

**Annexes**

**Annex 1**

Initial condition and flow in a dynamical system

A dynamical system is a set of differential equations, expressing the derivative of the current state from the current state (see for introductions Crawford, 1991; Strogatz, 1994). The system modeled evolves in a state space. The dynamical system gives a vector field acting upon the state space, which provides for any point in the state space its derivative, that is the instantaneous direction and size of its evolution. To get a trajectory, and the evolution thus time parameterization along it, one needs an initial condition somewhere in the state space, say X(0) at time t = t0, a time origin for this particular trajectory. The dynamical system defines a flow, that maps the initial state X0(t0) to a trajectory in the state space X(t). Dynamical systems have been designed to get unique solution from a given initial condition, whether the time is running forward or backward doesn't matter. If time is shifted, say by a time s like in X(t+s), the flow doesn't change, the same trajectory unfolds.

However, in a sense without a given initial condition there is no time origin separating past from future, the present. Even if often referred to, the present paper is not considering temporal processing per se, as involved for example in memorizing, reproducing, or comparing, temporal intervals. In that respect, neural circuits in many species have been shown to memorize time intervals over the range of tens of microseconds to hundreds of milliseconds, and more recently over the range of one to 20 seconds by rhythmic activities among specific areas (Sumbre et al., 2008). The extent to which and how the brain "can tell time" has been classically modeled as a single centralized clock relying on an oscillator and an accumulator (counter), or instead based on ongoing neural network dynamics (Karmarkar & Buonomano, 2007). The latter approach, the state dependent networks, is also used to model the emergence of a memory trace, which is captured by a physical analogy: "A useful analogy is the surface of a liquid. Even though this surface has no attractors (…) transient ripples on the surface can nevertheless encode information about past objects that were thrown in" (Ganguli et al., 2008; p 18970). Two important limitations are readily identified: "reliance on the state of a complex system to tell time creates potentially serious limitations due to the resulting dependence on the initial state and the lack of a linear metric of time" (Karmarkar & Buonomano, 2007; p 433). We proposed some embodied answer to such limits.

**Annex 2**

Example of code to generate the time series shown in Figure 5. The integration is performed using the Euler-Maruyama scheme (Higham, 2001). Parameters of the excitor dynamics are set to get a regime of one fixed point attractor and a slow manifold plus a separatrix (See Jirsa & Kelso, 2005). 12 « sounds » perturbations are added at times specified by da(i) 1 to 12, using the block function, basically a squared pulse.

The Matlab code Euler integration parameters calls the function single_excitator_JL which uses the function block to input discrete perturbations (step function) at times specified in d(i) = 1 : 12

```
% Euler integration parameters
dt = 0.03; % timestep
iters = 10000; %path length
time = (0:dt:iters*dt);
% Model parameters
a = 0.5;
b = 1;
c = 3;
```

```matlab
gamma = 1;
% Time points of perturbations
dt1 = 10; dt2 = 30; dt3 = 50; dt4 = 70; dt5 = 90; dt6 = 110;
dt7 = 130; dt8 = 150; dt9 = 170; dt10 = 190; dt11 = 210; dt12 = 230;
% Initial conditions
yin = (2 -0.8);
% Noise strength of the Langevin equation (see stochastic differential eq.)
D = (0.08 0);
y = yin;
 for ii = 1:length(time) % integrate over time
            ydot = single_excitator_JL(time(ii),y(ii,:),a,b,c,gamma,dt1,dt2,dt3,...
            dt4,dt5,dt6,dt7,dt8,dt9,dt10,dt11,dt12);
             y(ii+1,:) = y(ii,:) + dt*ydot + sqrt(2*D).*randn(size(ydot)).*dt;
    end

figure(1)
subplot(3,1,1);
plot(y(:,1),y(:,2),'g')
xlabel('x')
ylabel('y')
title('Excitator with noise')
subplot(3,1,2);
plot(y(:,1),'c.')
ylabel('x')
xlabel('time')
subplot(3,1,3);
plot(y(:,2),'k.')
ylabel('y')
xlabel('time')

function ydot = single_excitator_JL(t,y,a,b,c,gamma,dt1,dt2,dt3,...
         dt4,dt5,dt6,dt7,dt8,dt9,dt10,dt11,dt12)
duration_pert = 1;% in number of samples
da1 = block(t,dt1,duration_pert);   da2 = block(t,dt2,duration_pert) ;
da3 = block(t,dt3,duration_pert);   da4 = block(t,dt4,duration_pert);
da5 = block(t,dt5,duration_pert);   da6 = block(t,dt6,duration_pert);
da7 = block(t,dt7,duration_pert);   da8 = block(t,dt8,duration_pert);
da9 = block(t,dt9,duration_pert);   da10 = block(t,dt10,duration_pert);
da11 = block(t,dt11,duration_pert);   da12 = block(t,dt12,duration_pert);
   da = da1 + da2 + da3 + da4 + da5 + da6 +da7...
       + da8 + da9 + da10 + da11 + da12;
% excitator ODE
   ydot(1) = c*(y(2)+gamma*y(1)-y(1)^3/3 ) ;
   ydot(2) =  -(y(1)-a + da + b*y(2) )/c ;

function func = block(x,a,b)
%   equals one if a < x < a+b,
%   else zero
func = 0.5*(sign(x-a)-sign(x-(a+b)));
index_1 = min(find(func));
index_2 = max(find(func));
func(index_1) = 1;
```

func(index_2) = 1;

**References**

Banerjee, A., Jirsa, V. K. (2007). How do neural connectivity and time delays influence bimanual coordination?. Biological cybernetics, 96, 265-278.

Beuter, A., Bélair, J., Labrie, C. (1993). Feedback and delays in neurological diseases: A modeling study using dynamical systems. Bulletin of mathematical biology, 55, 525-541.

Binetti, N., Siegler, I. A., Bueti, D., Doricchi, F. (2010). Time in motion: Effects of whole-body rotatory accelerations on timekeeping processes. Neuropsychologia, 48(6), 1842-1852.

Capelli, A., Deborne, R., Israël, I. (2007). Temporal intervals production during passive self-motion in darkness. Current psychology letters. Behaviour, brain & cognition, (22, Vol. 2, 2007).

Cappe, C., Morel, A., Barone, P., Rouiller, E. M. (2009). The thalamocortical projection systems in primate: an anatomical support for multisensory and sensorimotor interplay. Cerebral cortex, 19, 2025-2037.

Chen, Y., Ding, M., Kelso, J. S. (1997). Long memory processes (1/f α type) in human coordination. Physical Review Letters, 79, 4501.

Clark, A. (1998). Being there: Putting brain, body, and world together again. MIT press.

Collins, J. J., Stewart, I. (1994). A group-theoretic approach to rings of coupled biological oscillators. Biological cybernetics, 71, 95-103.

Conte, A., Khan, N., Defazio, G., Rothwell, J. C., Berardelli, A. (2013). Pathophysiology of somatosensory abnormalities in Parkinson disease. Nature Reviews Neurology, 9, 687.

Crawford, J. D. (1991). Introduction to bifurcation theory. Reviews of Modern Physics, 63, 991.

Cullen, K. E. (2019). Vestibular processing during natural self-motion: implications for perception and action. Nature Reviews Neuroscience, 1.

Dallal, N. L., Yin, B., Nekovářová, T., Stuchlík, A., Meck, W. H. (2015). Impact of vestibular lesions on allocentric navigation and interval timing: the role of self-initiated motion in spatial-temporal integration. Timing & Time Perception, 3, 269-305.

Dieterich, M., Brandt, T. (2015). The bilateral central vestibular system: Its pathways, functions, and disorders. Ann. N. Y. Acad. Sci., 1343, 10–26.

Eagleman, D.M. (2004) The where and when of intention. Science, 303, 1144 –1146.

Eagleman, D. M., Peter, U. T., Buonomano, D., Janssen, P., Nobre, A. C., Holcombe, A. O. (2005). Time and the brain: how subjective time relates to neural time. Journal of Neuroscience, 25, 10369-10371.

Fitzpatrick, R.C, Day, B.L. (2004). Probing the human vestibular system with galvanic stimulation. J. Appl. Physiol. 96, 2301–16.

Ganguli, S., Huh, D., Sompolinsky, H. (2008). Memory traces in dynamical systems. Proceedings of the National Academy of Sciences, 105, 18970-18975.

Guerraz, M., Day, B.L. (2005). Expectation and the vestibular control of balance. J. Cogn. Neurosci. 17:463–469.

Haggard, P., Clark, S., Kalogeras, J. (2002) Voluntary action and conscious awareness. Nat Neurosci 5:382–385.

Higham, D.J. (2001). An algorithmic introduction to numerical simulation of stochastic differential equations. SIAM review, 43, 525-546.

Jirsa, V. K., Scott Kelso, J. A. (2005). The excitator as a minimal model for the coordination dynamics of discrete and rhythmic movement generation. Journal of motor behavior, 37, 35-51.

Jörges, B., López-Moliner, J. (2017). Gravity as a strong prior: implications for perception and action. Frontiers in human neuroscience, 11, 203.

Karnath, H.O., Dieterich, M. (2006). Spatial neglect - A vestibular disorder? Brain 129:293–305.

Karmarkar, U. R., Buonomano, D. V. (2007). Timing in the absence of clocks: encoding time in neural network states. Neuron, 53, 427-438.

Kelso, J. A. (1984). Phase transitions and critical behavior in human bimanual coordination. American Journal of Physiology-Regulatory, Integrative and Comparative Physiology, 246, R1000-R1004.

Kelso, J. S. (1995). Dynamic patterns: The self-organization of brain and behavior. MIT press.

Kelso, J. S., Fuchs, A. (2016). The coordination dynamics of mobile conjugate reinforcement. Biological cybernetics, 110, 41-53.

Lagarde, J., Kelso, J. A. S. (2006). Binding of movement, sound and touch: multimodal coordination dynamics. Experimental brain research, 173, 673-688.

Lee, D. N. (1976). A theory of visual control of braking based on information about time-to-collision. Perception, 5, 437-459.

Lopez, C., Schreyer, H.M., Preuss, N., Mast, F.W. (2012). Vestibular stimulation modifies the body schema. Neuropsychologia 50:1830–1837.

Milton, J. G., Cabrera, J. L., Ohira, T. (2008). Unstable dynamical systems: Delays, noise and control. Europhysics Letters, 83, 48001.

Monno, A., Chardenon, A., Temprado, J. J., Zanone, P. G., Laurent, M. (2000). Effects of attention on phase transitions between bimanual coordination patterns: a behavioral and cost analysis in humans. Neuroscience letters, 283, 93-96.

Moscatelli, A., Lacquaniti, F. (2011). The weight of time: gravitational force enhances discrimination of visual motion duration. Journal of Vision, 11, 5-5.

Salinas, E. (2006). How behavioral constraints may determine optimal sensory representations. PLoS biology, 4, e387.

Semjen, A., Leone, G., Lipshits, M. (1998). Temporal control and motor control: Two functional modules which may be influenced differently under microgravity. Human movement science, 17, 77-93.

Semjen, A., Leone, G., Lipshits, M. (1998). Motor timing under microgravity. Acta astronautica, 42, 303-321.

Seung, H. S. (1996). How the brain keeps the eyes still. Proceedings of the National Academy of Sciences, 93, 13339-13344.

Shampine, L. F., Thompson, S. (2001). Solving ddes in matlab. Applied Numerical Mathematics, 37, 441-458.

Sporns, O., Gally, J. A., Reeke, G. N., Edelman, G. M. (1989). Reentrant signaling among simulated neuronal groups leads to coherency in their oscillatory activity. Proceedings of the National Academy of Sciences, 86, 7265-7269.

Strogatz, S. H. (1994). Nonlinear Dynamics and Chaos with Student Solutions Manual: With Applications to Physics, Biology, Chemistry, and Engineering. CRC Press.

Sumbre, G., Muto, A., Baier, H., Poo, M. M. (2008). Entrained rhythmic activities of neuronal ensembles as perceptual memory of time interval. Nature, 456, 102.

Stetson, C., Cui, X., Montague, P.R., Eagleman, D.M. (2005). Illusory reversal of action and effect. Journal of Vision, 5, 769a.

Ramdani, S., Rossetto, B., Chua, L.O., Lozi, R. (2000). Slow manifolds of some chaotic systems with applications to laser systems. International Journal of Bifurcation and Chaos, 12, 2729-2744.

St George, R.J., Fitzpatrick, R.C. (2011). The sense of self-motion, orientation and balance explored by vestibular stimulation. J. Physiol. 589:807–13.

Schöner, G., Dose, M., Engels, C. (1995). Dynamics of behavior: Theory and applications for autonomous robot architectures. Robotics and autonomous systems, 16, 213-245.

Tass, P., Kurths, J., Rosenblum, M. G., Guasti, G., Hefter, H. (1996). Delay-induced transitions in visually guided movements. Physical Review E, 54, R2224.

Thakur, B., Mukherjee, A., Sen, A., Banerjee, A. (2016). A dynamical framework to relate perceptual variability with multisensory information processing. Scientific reports, 6, 31280.

Turvey, M. T., Shaw, R. E., Reed, E. S., Mace, W. M. (1981). Ecological laws of perceiving and acting: In reply to Fodor and Pylyshyn (1981). Cognition, 9, 237-304.

Varela, F. J., Thompson, E., Rosch, E. (1991). The embodied mind: Cognitive science and human experience. MIT press.

Venkadesan, M., Guckenheimer, J., Valero-Cuevas, F. J. (2007). Manipulating the edge of instability. Journal of biomechanics, 40, 1653-1661.

Warren, W. H. (2006). The dynamics of perception and action. Psychological review, 113, 358.

Zheng, C., Pikovsky, A. (2019). Stochastic bursting in unidirectionally delay-coupled noisy excitable systems. arXiv preprint arXiv:1902.06915.